# Redistributing Harm: Document Transition and the Limits of Trans Inclusion in India's Identity Systems

List of Authors:

Megh Marathe, Simon Fraser University
K Ranade, Tata Institute of Social Sciences
L. Ramakrishnan, SAATHII
Koyel Ghosh, Sappho for Equality

Contact address: megh_marathe@sfu.ca

## Abstract

Identity documents (IDs) are some of the most consequential sites through which transgender people encounter state infrastructures. This article examines document transition, the work of aligning names, gender markers, and related information across official records in India based on focus groups with sixteen trans participants, most of whom were transmasculine and gender diverse, in mid-2025. Participants described encountering an absence of clear protocols, changing demands for proof, and objectionable conduct from officials, arising often from limited understandings of transness in systems and policies. As a result, participants faced harms related to livelihood, housing, voting, travel, redress from violence and discrimination, and education. We situate these findings within information studies and trans studies scholarship on classification and state recognition. Participants with mismatching IDs faced a form of torque or administrative violence that we call 'trans tax' including disproportionate tax deductions, reluctant employers, and resultant insecure jobs. Further, gender-concordant IDs could redistribute harm: participants who became administratively legible as male lost jobs, benefits, and rights despite remaining vulnerable to the discrimination the programs sought to address. The disruption of people's lives by updated gender-concordant data calls for a re-examination of 'haunting' at the intersection of transness, classification, and data practices. The article concludes by calling for sensitization efforts directed at street-, system-, and policy-level decision-makers about trans identities and state provisions together with established and widely disseminated protocols for document transition in the short term; as well as a broad and coordinated upheaval of systems and policies to undo cisgender-heteronormative assumptions.

## Introduction

Identity documents are some of the most consequential sites through which transgender and gender diverse people (henceforth, trans people) encounter state information infrastructures. This article examines *document transition*, the work of aligning names, gender markers, and related information across official records, in India. Through focus groups with 16 trans individuals, most of whom were transmasculine and gender diverse, we demonstrate the limits of India's transgender identity card scheme.

Participants reported facing an absence of clear protocols, variations in practice, changing goal posts, and objectionable conduct by officials, arising often from limited understandings of transness in document transition processes and policies. As a result, participants faced harms related to livelihood, housing, voting, travel, redress from violence, and education. Importantly, harms did not end when participants obtained gender-concordant IDs. Instead, gender-concordant IDs generated new forms of exclusion: participants who became administratively legible as male became ineligible for certain jobs, benefits, and rights, despite remaining vulnerable to the discrimination the programs sought to address.

We situate these findings within information studies and trans studies scholarship on classification and administrative violence. Information studies has shown that people who do not easily fit into the representational schemes of state infrastructures face harmful life consequences (Bowker and Star, 1999), and may be held to higher evidentiary standards in bureaucratic interactions (Wyers, 2025). Our study nuances these insights. In addition to life harms and bureaucratic scrutiny, trans people with mismatching IDs faced what we call a 'trans tax' from both state and private entities. Limited understandings of transness in public and private backend data processing systems resulted in higher income tax deductions and an unwillingness to hire such people.

This trans tax is a form of "administrative violence" (Spade, 2011), a trans studies term used to describe harms perpetrated by state entities (Currah and Moore, 2008; Currah and Mulqueen, 2011; Nisar, 2018; Spade, 2011). Many harms may persist after updating IDs because old and inaccurate data continues to circulate in private corporations like credit bureaus, "maintaining a connection between a past and present identity" (Mackenzie, 2017: 45). Our study adds to this work, showing that for transmasculine people, *updated gender-concordant data* caused downstream harms and disrupted connections between past and present, resulting in lost jobs, benefits, and recognition for formal relationships. We propose the notion of haunting by gender-concordant data as a distinct phenomenon at the intersection of transness, classification, and data practices.

Our study focuses on document transition experiences under the 2019 Transgender Persons (Protection of Rights) Act (henceforth, TPPRA) and the associated 2020 Rules (Government of India, 2019; Ministry of Social Justice and Empowerment, 2020). Among other measures, the TPPRA laid down rules for legal transition on official documents. Following prior scholarship, we understand gender transition as a multifaceted process that may include social, medical, and legal components.[1] Social transition involves changes in gender expression and recognition, including name, pronouns, and presentation. Medical transition involves gender-affirming healthcare, such as hormone therapy or surgery. Legal transition involves changing names and gender markers on state-recognized documents. While often treated as a part of legal transition, distinguishing *document transition* enables an examination of the procedural dimensions of the work of aligning identity data and the material consequences of (in)congruent data. Our study sheds light on the hurdles document transition poses for Indian trans individuals despite over a decade of formal legal recognition.

The article is organized as follows: after reviewing related work, we provide brief background on the status of trans rights in India and describe the methods and data. The results sections describe on-the-ground document transition challenges followed by their systemic underpinnings and consequences. The discussion situates the article's contributions within scholarship on classification and administrative violence. The conclusion suggests practical implications for how some of these harms might be addressed, broadly calling for an undoing of cisgender-heteronormative assumptions in systems and practices at every level.

[1] Social, medical, and legal components of gender transition may not be experienced as linear, uniform, or essential by all trans people.

## Related Work

Classification and state infrastructures are known to have uneven impacts, advancing the interests of some while actively harming others. People who cannot be easily represented in an infrastructure's classification schemes tend to face great harm from the organizational processes that they support and sustain (Bowker and Star, 1999; Dourish, 2017). Bowker and Star describe these experiences through the concept of *torque,* or "the twisting that occurs when a formal classification system is mismatched with an individual's biographical trajectory" (Bowker and Star, 1999: 231). Drawing on cases from governance and medicine, they show that people can be severely harmed when formal systems do not accommodate their circumstances or operate inconsistently.

These concerns have been central to trans studies scholarship on state recognition. Spade develops the notion of *administrative violence* to describe harms faced by trans people whose IDs are not gender-concordant or are mismatched across agencies (Spade, 2011). Based on an examination of law and policy in the United States, Spade argues that these problems emerge because state agencies rely on inconsistent and discretionary evidentiary standards for recognizing gender transition. The resulting inconsistencies make it difficult for trans people to find housing, get jobs, travel, and vote. In the Indian context, Wyers conducted ethnographic fieldwork during the initial implementation of the transgender category, finding that "street level bureaucrats" such as doctors and welfare officials tend to operate based on mainstream representations of transness in religious mythologies and sociocultural narratives to decide who is a legitimate trans applicant (Wyers, 2024). They tend to believe that legitimate trans people are either intersex or belong to indigenous trans identities such as Hijra, Kinnar, etc. Individuals who do not conform to these stereotypes such as transmasculine and gender diverse people may be viewed with suspicion, as claiming trans status to fraudulently gain access to welfare schemes, and may consequently face a higher burden of proof (Wyers, 2025). She argues that bureaucratic understandings emerge from a combination of historical narratives, biologically essentialist beliefs, and mistrust towards unfamiliar communities, that sensitization programs should address.

Mackenzie highlights the temporal persistence of identity data. Name and gender changes in government IDs do not erase records maintained by private agencies. Based on a study of trans people's experiences with financial credit bureaus in the US (Mackenzie, 2017), he shows that outdated and inaccurate data remain attached to credit reports and background checks long after document transition, resurfacing trans histories in undesirable ways.

Building on this scholarship, we find that document transition in India is difficult due to implementation and systemic challenges, causing many harms. Further, updated gender-concordant data too can perpetuate harm in cis-heteronormative systems and institutions.

## Background

There is a long tradition of scholarship on the history of transgender rights in the Indian sub-continent (Banerjie, 2018; Hinchy, 2019; Saria, 2019; Vanita and Kidwai, 2006). Despite a history of criminalisation and social exclusion, trans communities have campaigned to seek political participation and voting rights in 1990s (Gayathri and Karthikeyan, 2016), passports issued under 'Other' category in 2005, enumeration as 'Other' gender in the Census (Office of the Registrar General & Census Commissioner, 2011), and the right to self-determined gender in a Supreme Court judgement in 2014 (*National Legal Services Authority (NALSA) v. Union of India*, 2014). The Indian parliament passed the TPPRA in 2019. The Ministry of Social Justice and Empowerment setup a national trans ID card, meant to serve as the basis for changing other official documents and a mechanism to access welfare schemes (Ministry of Social Justice and Empowerment, 2024). In our study, trans individuals sought to update their Aadhaar card (a unique national identity number), PAN (permanent account number used for taxation), driver’s license, educational certificates, passports, and voter IDs.

In 2026 the TPPRA was amended through substantial changes to the definition of the category ‘transgender’ (Ministry of Social Justice and Empowerment, 2026b). This article is based on data collected in 2025, and so we are unable to engage with the amendments. The government purportedly introduced these amendments to ensure that TPPRA benefits reach legitimate trans people and curb misuse. Notably, the government has neither provided data on misuse nor is the data of use by trans persons encouraging. The government issued approximately 33,000 national trans ID cards over a period of six years, a mere 8.3% of the number of trans persons enumerated in the 2011 Census (Ministry of Social Justice and Empowerment, 2026a). Government expenditure on schemes related to transgender welfare hovered between 0.4% to 9.55% of allocated funds from 2021-2024 (Das, 2024). It could be argued that the amendments are rooted in the ambivalence and inability of state machinery to hold diverse realities of trans communities using one restrictive and governable category.

## Methods

We conducted a qualitative study in two large metropolitan cities, one in Tamil Nadu and one in West Bengal. Participants were recruited through three LGBTQ+ serving organizations in these cities. The study was approved by our institution’s research ethics board.

**Data collection**. We conducted 5 focus group discussions with 16 trans individuals in April and May 2025. Focus groups were structured to elicit narrative accounts of participants’ experiences in seeking gender-concordant IDs, challenges they encountered in doing so, and factors that contributed to and helped navigate these challenges (Kvale, 2007). Focus groups lasted between 60 to 180 minutes each and were conducted at LGBTQ+ serving organizations’ offices. After obtaining verbal consent from all participants, the research team took notes in the form of jottings. Jottings were converted into detailed long form fieldnotes after each session (Emerson et al., 2011).

**Data analysis**. Fieldnotes were analyzed through a bottom-up inductive process. The research team wrote analytic memos (Emerson et al., 2011) and used descriptive codes (Saldaña, 2012) to identify emergent topics. The research team met to discuss memos, compare descriptive codes, and develop a cohesive code set. We identified three codes explicated in the paper: ground-level challenges, systemic problems, and life impacts.

**Participant demographics**. We report demographics in broad terms in the interest of participant safety and confidentiality. Participants were between 22 to 50 years old and represented a range of trans identities (8 transmasculine people, 4 nonbinary or gender diverse people, and 3 transfeminine people including sociocultural identities such as Thirunangai and Hijra), incomes (including community organizers, hourly-wage blue-collar workers, white-collar workers, and people with generational wealth), and linguistic backgrounds (including Tamil, Bengali, Hindi, and English). At the time of the study, participants resided in two metropolitan cities in Tamil Nadu and West Bengal; many had migrated from rural towns. We could recruit transfeminine individuals in one city, whereas transmasculine individuals participated in both cities.

**Positionality**. We are a team of trans and queer academics and community organizers with over fifteen years of experience working with trans and LGBTQ+ people in India. We understand that our findings will be filtered through our identities and experiences. Our positionality grants us unique insight into the subject matter, but it also creates the responsibility to reflect on how our identities shape the research process. We approach this study with an understanding of our caste and class privileges, including our access to educational and professional opportunities. We recognize that our experiences do not encompass the full range of experiences of people of different gender identities and socioeconomic backgrounds. These reflections informed all stages of this work including study design, analysis, and reporting.

## Results

We found that document transition challenges arose from implementation issues and structural problems, leading to many negative impacts on trans people's lives.

### Document Transition Under TPPRA: Ground-Level Experiences

Despite the 2019-2020 passage of the TPPRA and associated rules, document transition remained an uphill task. Ideally, it should be a smooth and quick process with the national trans ID serving as the basis for changing all documents (Ministry of Social Justice and Empowerment, 2024). We found that this imaginary did not play out on the ground due to absence of clear protocols, variations in practice, changing goal posts, lengthy procedures, expected continuities in the life trajectories of trans people, limited awareness, and objectionable conduct on the part of officials handling document transition processes.

First, participants reported that different government agencies and different officials in the same agency handled document updates in divergent ways. Community organizers such as Jacob observed variability across passport officials: *"the Madras High Court has ruled that proof of surgery is not needed, but regional passport officers decide that you need a surgery certificate. There is no standard operating procedure."* Similarly, Miko said that some passport officials asked for proof of complete medical transition: *"a lot of people are being refused a binary marker on their passport, because officials ask for proof of having had 'all' surgery."* These demands were in contradiction to guidelines from India's Ministry of External Affairs, which administers passport related matters (Ministry of External Affairs, 2022).

Such variations were also encountered at Aadhaar Seva Kendras (ASKs), i.e., service centres for Aadhaar cards. Aadhaar guidelines permit one gender change without any proof and additional changes based on national trans IDs (Unique Identification Authority of India, 2021). Despite explicit guidelines, community organizers Lalitha and Tejas found a lot of variation on the ground: *"some Aadhaar departments ask for surgery certificate, while others do not."* For example, Naveen was asked for proof of surgery despite providing a Tamil Nadu state trans ID[2] affirming his identity as a trans man: *"ASK people said we can give a T, not M, marker."* Indeed, different officials in the same ASK could have conflicting opinions, as Inder found to his dismay: *"I went to the ASK in person and asked them, 'here are the documents I have. Will I be able to change my information?' The officer said, 'yes, you can change everything.' That same day, I went back with the hard copies of those documents, and they refused to accept my application saying that it would get rejected."* These negotiations were inflected by dimensions of applicant identity: *"a person's class, caste, clothing, appearance, and ability to speak matter"* (Uchit). The support of a reputed figure could influence how officials worked: *"if you go with a powerful trans woman who knows how to negotiate with officials, the Aadhaar office relents and gives an M in people's cards. That is what I had to do"* (Naveen). Echoing this sentiment, Uma noted that some community organizations *"accompany trans individuals to help change their documents."*

[2] The state of Tamil Nadu's Social Welfare Department issues their own trans ID separate from the national trans ID. The state trans ID denotes individuals as trans man, trans woman, or intersex, and is required by welfare schemes run by the state government. National trans IDs denote individuals as male, female, and transgender, and are required by nationally administered schemes and documents.

Lacking such support, the burden of proof shifted repeatedly, with requirements increasing or changing across visits. Participants were given the runaround or turned away despite sufficient supporting documentation. Community organizer Uma outlined these responses as follows: *"Aadhaar officials will ask for a card, such as PF* (Provident Fund[3]) *card. Once the person brings that, they are sent back for another card, and so on in circles to get one document at a time as a delaying tactic."* The process was experienced as inconsistent, with rules changing across visits and applicants. Karan noted, *"even if a person has a lot of documentation per the Trans Act such as trans ID, gazette, newspaper ads, medical affidavit, and so on, local ASKs turn them away. The government-run ASK uses a mix of genuine reasons, unreasonable asks, and excuses to delay or deny people. One person was rejected because their surgery certificate had the doctor's signature but not stamp. When he got that, he was rejected because it did not have his photograph. Which surgical document has the patient's photo?"* Rahul's application was rejected due to a technicality: *"my notarized affidavit had five bullet points whereas their format had four points. Instead of ignoring [the extra point], they used it as a basis for rejection."* Naveen encountered trans ID-related excuses: *"if you only have national trans ID, they ask you to get the state trans ID. When I took the state ID, they asked for surgery proof despite my ID stating trans man."* Several participants were unable to update their Aadhaar, effectively barring document transition altogether.

Obtaining proof involved many time-consuming procedures, each with barriers and delays. For example, national trans ID card applications require a gender dysphoria (GD) certificate that can take several months to obtain depending upon the assessment process followed at the Psychiatry department of the local government institution. In the absence of standard assessment protocols for gender dysphoria or incongruence, the time taken as well as costs involved in getting the GD certificate vary (Ranade et al., 2025). As a result, even successful applicants waited over two months to obtain the national trans ID. The trans ID was only the first step of a long process involving a two-step Aadhaar update (*"change phone, address, and name first. Once that goes through, change photo and gender,"* per Naveen), followed by changes to other documents including *"family ration card (this can take 3 months), voter ID, PAN card (15-20 days), bank passbook, passport, etc.,"* per Lalitha. Every additional proof requested set timelines back by weeks or months. Participants reported many such instances, including Rahul: *"10-15 days after submitting my Aadhaar application, the portal said, 'Proof of identity is rejected'. The ASK officer said, 'you should wait a minimum of 30 days before [resubmitting the application]. After that period, a regional ASK official verified and approved all my documents [and uploaded them to the portal]. 30 days later, rejected again."* Eventually, he was unable to get his Aadhaar gender marker updated. That is, delays did not always resolve favourably.

Next, participants noted that document transition imposed financial burdens. Inder said, *"it quickly becomes expensive. The state gazette charges Rs. 2000-3000 and national gazette charges Rs. 10000-15000 [including advertisement and lawyer fees]. You must provide both: one for name change and another for gender change."* Rahul stated, *"an online law firm asked for Rs. 12000 for a direct change to M, but they were unable to make it happen."* After repeated delays and denials through official channels, some turned to private agents *"out of frustration and desperation"* (Chitrangada). Rahul noted that agents too were costly: *"there is no set cost. It depends on who refers you or how you find them. They can charge Rs. 5000, 7000, 10000."* These costs were not affordable for many.

Further, document transition procedures incorrectly assumed continuities in trans people's life trajectories including acceptance and support of birth families. Several participants had run away from hostile birth families and could not visit natal localities for safety reasons. Yet, official processes relied on documents such as school and college Bonafides that could only be obtained through those means. For instance, Naveen had to visit his native place to update his bachelor's degree certificate despite *"not being in touch with parents and not planning to."* Selvi's passport application was rejected at the police verification stage because she could not prove her educational credentials: *"my family burnt my marksheets [after I came out]."* There was no way to obtain duplicate marksheets due to having fled her native place for safety reasons. She reapplied for a passport declaring no formal education, but police verification was again denied, with officials stating that their system still showed the high school education from her previous application. Thus, assumptions of continued access to birth family and place limited avenues for document transition.

[3] Provident fund is another name for mandatory pension funds provided administered by the Employees' Provident Fund Organization, a social security agency under the Government of India.

Uninformed assumptions also showed up in relation to the national trans ID as a document. Participants found that officials possessed limited awareness around trans IDs, even in agencies that ostensibly accepted the trans ID as a proof of identity. For example, officials at a major government-run ASK questioned the authenticity of Yash's trans ID because its format and appearance did not match their expectations of an ID card. Yash explained, *"the trans ID is given as an electronic PDF. The PDF is pink, but the printout looks black & white, so they said this is fake. Also, it is not sized like a PAN or Aadhaar card, so they turned me away."* Another official asked Karan to, *"bring me the original."* He tried advocating for himself, saying, *"sir, there is no original. This is it. It's a PDF."* The official remained unconvinced and claimed that an earlier applicant had a trans ID that *"looked more like a regular card."* Karan's best explanation was that someone had resized the PDF down to the size of a regular ID card, taken a colour printout, and laminated it to make it seem official. Police officials too seemed unaware, noted community organizer Miko: *"a 20-year-old trans man was held captive by family. When we complained in the police station, the officer in charge asked for proof. We gave him the trans ID, but he insisted it is fake and that no such card existed,"* thereby impeding their efforts to secure justice. Educational institutions also seemed uncomfortable using the trans ID as the base document for updates. Despite providing a trans ID, Jacob was asked to submit an affidavit signed by their school principal juxtaposing their birth name and old photo with their chosen name and recent photo to update their 10$^{th}$ and 12$^{th}$ standard marksheets. Participants urged for more awareness around the trans ID, including its digital format and validity as identity proof.

Finally, participants faced insensitive behaviour from officials both while obtaining a national trans ID and later using it to update other documents. Miko spoke about the plight of trans ID applicants: *"at the district magistrate's office, people often get asked very insensitive questions and treated badly. Especially transmasculine people and those living in the outskirts. [Officials] ask irrelevant and intrusive questions like, 'can you have a baby?' or 'what surgery have you had?' People's self-esteem is crushed. Showing up with old documents adds to it. We offered to conduct free sensitization programs, but there is no interest."* Yash shared a traumatic experience at a government-run ASK: *"I had a horrible experience. The first person who saw my documents had no idea what a transperson was or what the procedure was. I was asked all kind of questions and kept waiting for 2-3 hours while my case was shunted all over the office. I could see people laughing, making fun, and discussing it among themselves."* Yash and others were forced to tolerate demeaning and disrespectful conduct. Objecting to this behaviour was not seen as a possibility. Instead, participants *"made peace with the fact that polite behaviour cannot be expected from officials"* when seeking identity-concordant documentation (Rahul).

In summary, gender-concordant documentation seemed a distant and even insurmountable goal due to shifting and inconsistent criteria, delaying tactics, time-consuming processes, natal family reliance, limited awareness, and insensitive treatment.

## Systemic Challenges as Roadblocks for Realization of Trans Rights

We found that systemic problems including limited understandings of transness compounded the implementation challenges described earlier.

The clearest example of a systemic problem came from Tamil Nadu, which participants described as, otherwise, one of the most trans inclusive states from a policy and implementation standpoint. Tamil Nadu established a transgender welfare board in 2008; the first Indian state to do so (Pramod and Jeganathan, 2025). The board was originally named the Aravanigal Welfare Board, referring to transfeminine and gender-variant people who worship Lord Aravan, a deity. It was later renamed to the Thirunangai Welfare Board, with thirunangai being a compound of ‘thiru’, the equivalent of sir, and ‘nangai’, meaning woman in Tamil, thus standing for *“honourable woman and also male woman,”* per community organizer Tejas. The term thirunangai gained popularity in political and government circles, soon becoming *“the term used everywhere,”* even after the board was briefly renamed the Transgender Welfare Board due to community pushback. Tejas explained, *“the emphasis on thirunangai is a big problem because the government insists on using the term for all transpeople, even trans men. For example, we are now back to the Thirunangai Welfare Board. Transmasculine people would like to use the term thirunambi, because ‘nambi’ means man.”* Although the board website now uses more inclusive terms such as ‘thirunar’ (meaning trans person), its initiatives continue to employ the term thirunangai: for instance, the mobile app used to obtain the state trans ID and access state-run welfare schemes is titled the Thirunangaikal (trans women’s) mobile app (Tamil Nadu Social Welfare and Women Empowerment Department, 2025). The board also had tokenistic transmasculine representation, which was a problem because *“transmasculine people’s access to welfare schemes depends on how good their equation is with local thirunangai and jamat leaders”* (Jacob). That is, transmasculine people remain outside the social, political, and bureaucratic imaginaries of transness, needing to affiliate with transfeminine nomenclature and groups to access state-sanctioned resources.

Potentially as a result, participants reported that transmasculine people were asked to meet a higher burden of proof and/or treated worse during document transition. In community organizer Lalitha’s experience, ASK officials asked transfeminine people for proof of “*any one surgery, including [facial] feminization surgery”* whereas trans men were asked for proof of lower surgery. Similarly, Miko noted that passport officials asked transmasculine people for *“complete surgery documentation.”* Officials also tended to respond better to trans women and those in influential positions. Jacob said, *“if you go with a powerful trans woman from [community organization] then it’s easy. If many trans guys go together, one of whom is on the Trans Welfare Board, it’s possible. If a single trans guy goes, they will definitely discriminate.”* As mentioned earlier, Naveen’s Aadhaar gender was updated to M only after being accompanied by a trans woman community organizer. Some held that *“it is much easier for transfeminine folks to get their documents changed”* (Tejas). Others added that transmasculine people were treated worse: *“trans women are not better off in terms of ID change, but transmasculine people are more infantilized by officials”* (Palash). Lalitha speculated that this difference arose because *“officials are afraid of trans women – they think trans women have the power to curse – or they [sexualize trans women and] are attracted to them.”* Others added that both fear and sociocultural understandings likely played on officials’ minds and helped in the case of trans woman applicants.

Limited understandings of transmasculine identities were also reflected in welfare policies and laws, determining who could access ration cards, health insurance, pension, and redress from law-and-order agencies. Single trans men with updated IDs were not eligible for family ration cards while trans women were, in accordance with cisgender standards, explained Jacob: *“single trans guys can’t get a family ration card because single cis men can’t, while single trans women can because single cis women can. Even if they get one, the* *ration card shows thirunangai (trans woman) for trans men."* Welfare schemes were tied to ration cards, so trans men could not access state health insurance for gender-affirming care: *"there is a need to de-link the state insurance scheme from ration cards. Trans guys struggle to access state health insurance and gender-affirming care [without ration cards]."* Along similar lines, Miko stated, *"an AFAB transmasculine person was getting their parents' pension in their deadname.*[4] *Now, they are not getting it [after updating documents]."* That is, gender-concordant documentation put this person's financial wellbeing in peril. Narrow understandings also imperilled transmasculine people's ability to seek justice in case of crimes or allegations against them. This data is presented in the next section; in essence, interpersonal violence tended to be framed as violence against women by men or between men. Both framings placed transmasculine people in precarious positions.

Peculiarities in major identity systems' demographic fields further attested to limited bureaucratic imaginaries of transness. We focus here on Aadhaar, PAN, and passports. The Aadhaar system has a care-of field whose value is inferred from the card-holder's gender in all cases except legal guardianship, where a gender-neutral "C/O" value is used. The care-of field precedes the address printed on the physical card. That is, addresses commonly begin with gendered relational indicators such as "D/O [daughter of] Mohd Shaikh" and "S/O [son of] Rama Devi."[5] Changing the gender marker in the Aadhaar system did not update this relational indicator. That is, trans men's cards retained "D/O" and trans women's said "S/O," despite opposing gender markers in the same document. Karan shared the example of a friend who faced this issue: *"he received a new card with his chosen name, gender, photo, and address. But his card still says daughter of instead of son of."* There was no option to update the care-of field or even change it to the gender-neutral value. That is, Aadhaar policymakers and system developers seemed to hold that gender data updates would be restricted to gender alone, despite the presence of other inferred fields.

Further, Aadhaar name change rules seemed to be designed without consideration for minoritized religions, in that trans applicants were only allowed to update their first name. Community organizer Chitrangada highlighted the problems with this approach: *"Aadhaar name changes are problematic for people from non-majority religions and backgrounds. Last names can be gendered for Muslim and Sikh people: Khatoun for AFABs and Sheikh for AMABs and similarly Kaur and Singh. But officials insist that you can only change first name, meaning that [updated] IDs continue to carry birth gender [signifying] names."* Such signifiers could complicate passing, negatively impacting people's personal and professional lives.

We speculate that narrow imaginaries of transness might also be why trans people had to resort to a two-step Aadhaar update strategy, changing name and address in the first step and gender and photo in a separate step. Participants across states said this strategy was proven to minimize rejection: if applicants requested all changes at once, they would be turned down. The notion that a trans person updating their gender would likely want to update their name and photo and may indeed need to change their address to escape a hostile family situation appeared to have escaped the bureaucratic imaginary.

The PAN system projected a veneer of inclusion while leaving at least one key field unchanged, leading to non-consensual disclosure and outing in the workplace. Naveen had updated the name and gender in his PAN records, and received a physical card displaying his chosen name, along with PAN number and father's name as is typical. The physical card does not contain an honorific or gender marker. During monthly backend processing, his colleagues found that the title in his PAN records was displayed as 'Shrimati' (Ms.) instead of 'Shri' (Mr) in the system. This discrepancy likely resulted either from a missing update to the inferred honorific field or due to data inconsistencies across linked systems (his PAN and Aadhaar records were linked as a prerequisite to receiving income, and he *"applied to change PAN before the second step of Aadhaar update [gender change] was complete,"* meaning that the gender listed in his Aadhaar record took precedence over the PAN record).

---

[4] 'Daughter[s]' are eligible to receive family pension for deceased parents in India (Comptroller and Auditor General of India, 2020).

[5] The care-of field is technically optional, but in practice is nearly always filled in. People can enroll in the Aadhaar system in two ways: by referencing the Aadhaar number of a family member (called the Head of Family route) or by independently submitting identity proof documents. The care-of field is mandatory for the Head of Family route and operators also tend to fill it in for the other route (Unique Identification Authority of India, 2025).

Passports posed another set of challenges due to in-built assumptions about the nature of family constituting of cisgendered heterosexual units. Jacob shared: *"my amicably divorced ex-husband wanted to apply for passports for our kids. The passport official showed us that there is no way to list two fathers either in the application form or the physical passport. The only fields are father's name and mother's name. He showed us that there is no flexibility; the field names are hardwired in the system."* Jacob's workaround was to leave the mother's name field blank and prefix the open-text address field with *"C/O Jacob."* Ultimately, however, Jacob's children's passports do not recognize him as a legitimate parent despite the children living with him as their primary caregiver.

Bureaucratic imaginaries also tended to exceptionalize trans people's experiences. For instance, Tamil Nadu trans IDs used the term *"third gender name"* (Jacob) to designate chosen name, as though it were an entity entirely different from a cisgender person's name.

Together, problems in bureaucratic systems and policies complicated document transition for all and more so for transmasculine people and those from non-majority backgrounds.

## Life Impacts

The slow and even insurmountable process of document transition produced problems related to livelihood, housing, voting, travel, redress from violence and discrimination, and education.

Livelihood impacts arose from counterintuitive hiring practices and mismatches across identity documents and/or physical appearance. To begin, large firms required trans people hired under their diversity, equity, and inclusion (DEI) quotas to update identity documents, but the very process and outcome of such updates could place their livelihoods at risk.

Naveen's company gave him two months to update his name and gender on Aadhaar and PAN cards. He began with Aadhaar updates using the two-step process known within the community. The first step took one month. In the second step, his binary gender change request was rejected, and he was asked to provide proof of surgery, which he did not yet have. The PAN update took well over two months. Since he could not update documents within the required timeframe, the company terminated his employment. He was hired by another big firm under their DEI quota. His employer-provided insurance partly covered top surgery, helping Naveen obtain a surgery certificate. Since the company allowed limited leave, typically two days per month, he used the post-surgery recovery period to complete document transition procedures. When he returned to work after surgery, his documents reflected a male marker. However, his employer said they could not retain him: *"they said, 'we need documents with a T marker, otherwise the backend system will complain.'"* Community organizer Tejas explained, *"companies don't want to show you as M. They prefer to show new hires [using] T or F [markers]. An M goes against their DEI goals of hiring more women."* That is, only trans people with T or F marker documents could retain their jobs. No alternative was offered, and Naveen was forced to look for a different job.

Participants who had begun, yet not completed, document transition encountered problems across linked systems. Given the varying timelines and requirements of different agencies, participants sometimes had inconsistent records, such as chosen name and gender in PAN records but chosen name and assigned gender in Aadhaar records. Such discrepancies resulted in direct and indirect penalties. During monthly at-source tax deduction (TDS) processes, the national employee provident fund (EPF) system would incorrectly deduct a *"flat 20% TDS, which is much higher than the usual rate of ~10% salary-based tax"* for employees whose data did not precisely match across PAN, Aadhaar, and EPF records (Tejas). Ordinarily, people could apply online to change their information in the EPF system pending employer approval. Yet, the EPF system rejected binary gender changes even after employer approval, providing the following reason: *"a new person is not allowed to hold the same EPF number"* (Uma). That is, the system failed to recognize the possibility of binary gender transition. The EPF system also displayed an error if an employee began document transition and changed, say, the name on their PAN but not their Aadhaar: *"although the [PAN and Aadhaar] numbers remain the same, the system shows a pop-up saying name mismatch" (Uma).* This error occurred even though both unique identifiers (i.e., PAN and Aadhaar numbers) remained the same. Employers could not help trans individuals in either case. As a result, many companies refused to hire trans people who did not have matching documents, reported Uma: *"companies ask trans people to get Aadhaar, PAN, and EPF to have the same information, otherwise they do not hire."*

As a result, trans people were pushed to insecure, hourly wage, or gig economy jobs that required fewer documents but excluded them from employee pension and insurance benefits: *"people end up joining Swiggy, Zomato, Porter [delivery services] who don't give PF. Or they drive electric bikes and rickshaws that don't need driver's licenses"* (Uma). Many such jobs required driving but the ease of updating driver's licenses varied across states. While Tamil Nadu made it easy to update licenses using state trans IDs *("apply online using gazette, trans ID card, and old driver's license. They will call you in person to verify documents and sign. Within 30 minutes, the update happens. Soon, one can get a soft copy and hard copy follows later,"* per Uma), no West Bengal participant was able to update their license. For people who had transitioned medically and socially, this led to an obvious mismatch between their license and physical appearance. Yash said he always carried his old Aadhaar card (despite having updated it) in case of a traffic police stop: *"I show them my old Aadhaar card and driving license so that the cops understand my situation."* He reasoned that the matching assigned name and gender on his driver's license and old Aadhaar card would help clarify the mismatch between his license and appearance. While such workarounds facilitated trans people's livelihoods, they also necessitated disclosure and potentially placed them in high-risk situations.

The instability caused by mismatches between appearance and documentation extended to other realms including housing, voting, travel, and redress for discrimination or violence. Participants reported difficulties in finding stable housing: Inder and Karan rented an apartment when they first moved to the city several years ago. Both had since transitioned socially and medically and passed as cis men but had been unable to change IDs (due to challenges described in previous sections). Their landlord, a senior citizen, had not noticed and/or commented on their transition, perhaps because they were model tenants: *"she doesn't seem to notice so long as we pay rent on time"* (Inder). The landlord's children had recently started managing rentals. Inder and Karan were afraid of losing their flat once the new landlords noticed that *"there are now men in the apartment"* (Inder). They noted that it was *"impossible for trans people to sign new leases given the mismatch between appearance and documents"* (Karan). Although some state governments had announced schemes providing free homes to low-income trans ID card holders, community organizer Selvi reported limited implementation on the ground. Several participants held on to their current housing situation, hoping landlords did not discover or object to their transness.

Participants whose documents had not caught up with their appearance reported being afraid to vote and travel. Karan was worried about how people in his natal locality would react to his physical appearance (which differed considerably after medical transition). After many years of abstaining, he went to vote two years ago but *"at the very end, when the voting booth was almost closing, to make sure I wouldn't meet other people."* Other participants worried about a lack of awareness from election officials. Inder went to the voting booth with *"my old voter ID, Aadhaar card, and [current] trans ID, looking very masculine after 1.5 years on hormone therapy."* The discordance between his appearance and IDs caused confusion but he was eventually allowed to vote: *"people at the election booth looked confused but seemed to understand something and let me vote."* These narratives suggest that slow nature of document transition may impede trans people from voting. Several participants felt that this slowness additionally constrained their mobility. Inder said, *"it is impossible for me to travel for fun or work because all hotels ask for an ID and address proof [and mine do not match my appearance]."* As a result, *"I have not been on a trip for 3-4 years."* Thus, document transition issues prevented participants from traveling for work and going on vacation.

Documentation status also affected access to redress in case of crimes against transmasculine people. Manoj experienced physical violence from a politically connected man in public. Legal processes required him to *"give original ID,"* which he had been unable to update despite years of social and medical transition. His lawyer advised framing the case under the Indian Penal Code's Section 354 (Union of India, 1860), commonly referred to as *"outraging the modesty of a woman,"* requiring him to *"present as a woman and get deadnamed and misgendered throughout to seek justice"* (narrated Uchit). In a parallel scenario, community organizer Miko reported that a trans man was held in a male cell based on false allegations by his partner's unsupportive parents: *"he was booked under trafficking, theft, forgery, and kidnapping of his cis woman partner. [The police] kept him in a male cell."* In yet another case, they added, *"a married trans man got into trouble because his wife claimed that she did not know her partner was trans. He was booked for fraud and fraudulent documents."* In hourly wage jobs, Uchit said trans people had limited recourse in case of workplace discrimination such as *"10-11 hours of standing as a security guard with no bathroom break."* Participants described all situations as non-ideal and resulting from gaps in the TPPRA, which contained anti-discrimination and protection clauses without enforcement mechanisms ("*helplines for crisis cases and also for documentation don't exist,"* said Miko).

Finally, document transition problems complicated college admissions. Community organizer Uma said that for trans people who had changed one document such as Aadhaar but not others, *"there is a high chance that [they] may not get admission"* in postsecondary institutions. In her experience, *"school documents are the hardest to change."* Schools often refused to provide Bonafide certificates (documents attesting to an individual's school affiliation, which can serve as proof of identity) because *"the school or the current principal may not want to take responsibility."* Naveen was asked to provide a surgery certificate to change his 10$^{th}$ and 12$^{th}$ standard marksheets by the Department of Public Instruction. Even after doing so, it took them another six months to issue updated marksheets. Community organizer Jacob said, *"many people have still not gotten updated certificates over 18 months [after application],"* potentially limiting their right to postsecondary education.

In summary, document transition problems compromised participants' ability to make a living, find housing, vote, travel, seek justice, and pursue education.

## Discussion

This study finds that document transition processes under TPPRA fail to meet the needs of trans people, threatening their access to secure livelihood, housing, voting, mobility, redress from violence, and education. We discuss implications for information studies and trans studies conversations on torque, administrative violence, and sensitization work.

### Trans Tax, Private Individuals, and Haunting

We find that document transition processes torque the lives of trans people in line with scholarship on classification (Bowker and Star, 1999). We confirm Wyers's hypothesis that transmasculine and gender diverse people would face a high burden of proof due to being farther from bureaucratic imaginaries of transness (Wyers, 2025). In addition, we trace the consequences of high evidentiary standards, showing that they extend beyond bureaucracy and range from basic rights and necessities to leisure. Participants literally and metaphorically faced what we call a 'trans tax': e.g., tax assessments incommensurate with salary, employers' unwillingness to hire, and resultant jobs without pension or insurance. Trans taxation is a form of administrative violence (Spade, 2011), but it may be more widespread than commonly described. Prior work has documented harms perpetrated by the state such as in government offices, prison systems, shelters, and security mechanisms (Currah and Mulqueen, 2011; Nisar, 2018; Spade, 2011), with some attention to private firms (Mackenzie, 2017; Spade, 2011). We find that private individuals and establishments such as landlords, hotels, and educational institutions may play a major role in perpetuating violence. There is a need to investigate trans taxation in individual and private contexts.

Importantly, our study finds that transmasculine people face harms even after accurate representation in identity systems, in contrast to prior work. The literature theorizes that people face harms because they are not well-represented, with the implication that better representation may alleviate harms (D'Ignazio and Klein, 2020). Yet, several participants reported harms after, and notably due to, obtaining gender-concordant IDs: Naveen was fired after completing document transition as required by his employer, Miko's friend lost their pension, and Jacob was not documented as a parent on their children's passports. Harms occurred in both private and government settings.

Experiences of harm despite fitting in are underexamined in prior work. How should these cases be understood and theorized? We propose the notion of haunting. Despite meeting formal standards and correcting identity records, individuals experienced harms similar to their misfitting counterparts. Here, haunting arose due to poorly designed systems that did not account for trans people as whole individuals with identities, relations, and needs. Our notion of haunting is related to but distinct from Mackenzie's "afterlife of data": Mackenzie refers to how "outdated data stick, maintaining a connection between a past and present identity" (Mackenzie, 2017: 45). In contrast, in our case, up-to-date gender-concordant data caused negative effects. Further, updated data had the opposite effect, disrupting connections between past and present: people lost jobs they held, benefits they received, and relationships that were documented. Thus, in haunting, we describe a distinct phenomenon at the intersection of transness, classification, and data practices. The phenomenon was gendered in that transmasculine individuals seemed particularly impacted due to cisgender-heteronormative imaginaries underlying state welfare schemes and company DEI quotas, but this may also be an artifact of our dataset. There is a need for further research to develop a fuller understanding of haunting and its mode of operation.

Finally, some empirical observations in our study differed from prior work. Participants navigated hurdles using workarounds, which remain underexamined (Mackenzie, 2017). Examples included using creative interpretations to accommodate queer relationships, carrying old and new sets of IDs, and using private agents after repeated denials via official means. While participants' workarounds ranged in terms of the amount of risk and disclosure, they saw these as acceptable interim strategies in the face of unjust and discriminatory systems. Attention to workarounds in future research may shed light on ways to balance their risks and benefits. Administrative violence's impacts on people's ability to seek education and redress from violence also remain underexamined: educational impacts are commonly examined post admission (Kirks-Cler et al., 2025) and redress from violence has been examined primarily in terms of conditions in prison and correction systems (Spade, 2011). Our findings suggest the need for more research on how mismatching documents impact entry into educational systems and framing of legal disputes and arguments.

## Needs for Awareness, Protocols, and Sensitization on Document Transition

Participants reported encountering an absence of clear protocols, variations in practice, and changing goal posts in document transition processes under the TPPRA as well as limited awareness and objectionable conduct in various state agencies. We confirm Wyers's finding that bureaucrats were unfamiliar with trans identities that did not conform to sociocultural stereotypes (Wyers, 2024). Transmasculine and gender diverse participants reported especially insensitive treatment in government offices potentially due to this unfamiliarity. We agree with Wyers that officials should be trained to build familiarity with and overcome mistrust of trans communities via sensitization programs.

Additionally, in our study, bureaucrats were reportedly unaware of government provisions too. Officials tended to be unfamiliar with the national trans ID, its ability to serve as a proof of identity, its intended function as the basis for updating other official documents, as well as its digital format and appearance. Further, some bureaucrats asked for proof of (lower) surgery for gender updates in passports and Aadhaar cards, despite national guidelines and court rulings to the contrary. Trans participants with binary gender IDs were also disqualified from welfare schemes at times, again in the absence of formal requirements. These findings suggest that awareness and sensitization programs need to go beyond educating bureaucrats about trans people. Officials also need sensitization about the trans ID and government criteria for document transition and welfare schemes.

Participants experienced changing goal posts, where the evidentiary burden increased or changed repeatedly across visits to a single agency. We speculate that this response might arise from bureaucrats' attempts to navigate uncertainty created by the lack of clear protocols and limited awareness about trans identities and document transition. By turning applicants away despite sufficient documentation, officials perhaps sought to absolve themselves of the potential liability associated with approving updated IDs. This hypothesis draws on the insight that practitioners faced with uncertainty tend to adopt a risk-averse stance: for instance, Shuster found that medical practitioners passed their uncertainty onto trans patients in the case of gender-affirming care (Shuster, 2021). Sensitizing officials about document transition procedures may be one way to address this uncertainty.

Finally, bureaucrats and policies appeared to treat trans people as just another marginalized group in need of upliftment. That is, they expected that the common strategy of providing said group with access to welfare schemes mediated by some kind of ID (such as India's below-poverty-line cards) would suffice. The state and its bureaucrats were in a sense enacting a magnanimous gesture by providing a trans ID and welfare schemes. While many trans individuals need and deserve access to welfare, that this is all they need (or deserve) operationalizes a limited understanding of transness. The ideas that trans people may not affiliate with birth families, may seek passports for their children, and may desire accurate relational indicators on ID cards, instead of merely accepting whatever the state provides, did not seem to figure in the logics of identity systems and policies. This absence suggests the need for sensitization programs aimed at system developers and policymakers too.

We advocate for sensitization programs that educate a broad swathe of street-, system-, and policy-level decision-makers about trans identities as well as state and legal provisions, together with established and widely disseminated protocols for document transition.

## Conclusion

This study examined *document transition*, the work of aligning names, gender markers, and related identity information across official records, among trans people in India. Drawing on five focus groups with sixteen participants in mid-2025, it surfaced the limits of India's transgender ID scheme, especially for transmasculine and nonbinary individuals.

The findings highlighted several instances where document transition processes relied on cisgender-heteronormative assumptions that did not hold for many participants. Specifically, document updates seemed to target a small set of fields without considering fundamental assumptions undergirding systems and policies. As a result, participants did not fit into state apparatuses as a matter of routine. That is, insufficient consideration of trans people's circumstances and life trajectories rendered document transition inefficient *by design* (Bowker and Star, 1999; Dourish, 2017). Ongoing special intensive revisions of electoral rolls in states across India (Choudhury, 2025; Krishna and Rao, 2026) and recent amendments to the TPPRA (Ministry of Social Justice and Empowerment, 2026b) both impose additional evidentiary burdens based on similar assumptions and are likely to amplify these challenges and inefficiencies.

What might need to happen to remedy these inefficiencies? We hold that systems and policies need to be re-examined at every level, including inbuilt sociocultural assumptions and the scope of technical changes. To begin, there is a need to build in a comprehensive understanding of trans people's circumstances such as unsupportive or unsafe natal families and localities and the resulting need to break ties and migrate, often to urban centres, to facilitate survival. Evidentiary standards for document transition processes must be modified to accommodate these circumstances. Further, there is a need to recognize a diversity of trans identities including transmasculine and gender diverse people in designing mechanisms for welfare, law, and order. Cisgender-heteronormative standards are known to be a structural problem (Spade, 2011; Stryker, 2013) and need to be identified and potentially deprogrammed from a wide range of mechanisms including ration cards, pension schemes, and penal codes to afford protections for transmasculine individuals.

These wide-ranging impacts suggest the need for coordinated efforts across state agencies as well as public and private institutions that serve fundamental rights (such as education). Such infrastructural changes tend to be difficult to execute and require serious commitment in terms of coordination and governance across heterogeneous actors and organizations (Jones et al., 2023; Lee et al., 2006). Instead of tokenistic updates to a small number of fields, our study suggests the need for such a coordinated and systematic commitment.